# Scanning localized magnetic fields in a microfluidic device with a single nitrogen vacancy center

Kangmook Lim[a,b], Chad Ropp[a], Benjamin Shapiro[c], Jacob M. Taylor[b], Edo Waks[a,b,1]

[a]Department of Electrical and Computer Engineering and Institute for Research in Electronics and Applied Physics, University of Maryland, College Park, Maryland 20742, USA.

[b]Joint Quantum Institute, University of Maryland and the National Institute of Standards and Technology, College Park, Maryland 20742, USA.

[c]Fischell Department of Bioengineering and the Institute for Systems Research, University of Maryland, College Park, Maryland 20742, USA.

[1]To whom correspondence should be addressed. E-mail: edowaks@umd.edu.

**Abstract**

Nitrogen vacancy (NV) color centers in diamond have emerged as highly versatile optical emitters that exhibit room temperature spin properties. These characteristics make NV centers ideal for magnetometry, which plays an important role in chemical and biological sensing applications. The integration of NV magnetometers with microfluidic systems could enable the study of isolated chemical and biological samples in a fluid environment with high spatial resolution. Here we demonstrate a method to perform localized magnetometry with nanometer spatial precision using a single NV center in a microfluidic device. We manipulate a magnetic particle within a liquid environment using a combination of planar microfluidic flow control and vertical magnetic actuation to achieve 3-dimensional manipulation. A diamond nanocrystal containing a single NV center is deposited in the microfluidic channels and acts as a local magnetic field probe. We map out the magnetic field distribution of the magnetic particle by varying its position relative to the diamond nanocrystal and performing optically resolved electron spin resonance (ESR) measurements. We control the magnetic particle position with a 48 nm precision and attain a magnetic field sensitivity of 17.5 $\mu T\ Hz^{-1/2}$. These results open up the possibility for studying local magnetic properties of biological and chemical systems with high sensitivity in an integrated microfluidic platform.

**Significance Statement**

Nitrogen vacancy (NV) color centers in diamond enable local magnetic field sensing with high sensitivity by optical detection of electron spin resonance (ESR). The integration of this capability with microfluidic technology has a broad range of applications in chemical and biological sensing. We demonstrate a method to perform localized magnetometry in a microfluidic device with nanometer spatial precision. The device manipulates individual magnetic particles in three dimensions using a combination of flow control and magnetic actuation. We map out the local field distribution of the magnetic particle by manipulating it in the vicinity of a single NV center and optically detecting the induced Zeeman shift. Our results could open up the possibility for real-time sensing applications in a microfluidic platform.

**Introduction**

Nitrogen vacancy (NV) color centers in diamond possess remarkable properties that include single photon emission (1, 2), a spin-triplet ground state with long spin coherence time at room temperature (3, 4), and spin dependent photoluminescence (5, 6). These properties enable NV centers to act as highly sensitive magnetic field sensors with nanoscale spatial precision (7-11). NV magnetometry has been deployed in a variety of applications such as detection of nanomechanical oscillations (12), readout of magnetic data (13, 14), monitoring ion concentrations (15-17), and magnetic imaging (18, 19). NV centers in diamond are also bio-compatible (20-22) and can thus serve as biological sensors (23-25). This capability has motivated recent efforts to manipulate and control a diamond nanocrystal hosting an ensemble of NV centers (26) and a single NV center (27) in a liquid environment using optical trapping techniques.

The integration of NV magnetometry with microfluidic systems could open up new possibilities for real-time biological and chemical sensing. Microfluidics provides an ideal device platform for sorting and manipulating samples using magnetic, optical, mechanical, or electrical methods (28, 29). An important recent example is the work of Steinert et al. (30) that demonstrated a spin relaxometry technique based on an array of NV centers in a microfluidic device. This technique was able to measure the concentration of a large ensemble of freely diffusing, unperturbed magnetic ions and molecules. However, the microfluidic device in that work was only used to prepare appropriate ion concentrations. In addition to preparing fluid concentrations, microfluidics can also isolate and manipulate individual objects. For example, a variety of control methods based on electrokinetic (31-33) and magnetic tweezers (34-36) enable manipulation of objects inside a microfluidic device with nanometer spatial precision (37-39). These methods can enable localized magnetometry with high spatial resolution.

Here we demonstrate the ability to perform NV magnetometry on suspended objects in a microfluidic system with high sensitivity and with high spatial resolution. We manipulate a single magnetic particle in three dimensions using a combination of planar electroosmotic flow control and vertical magnetic actuation. A diamond nanocrystal containing a single NV color center acts as a localized magnetic field sensor for the manipulated magnetic particle. We measure the magnetic field generated by the magnetic particle via optically detected electron spin resonance (ESR) measurements, and map out its field distribution. Our results open up the possibility for highly sensitive magnetic field measurements in an integrated microfluidic system.

**Experimental Procedure**

Figure 1a illustrates the microfluidic system we have developed to manipulate magnetic particles and perform localized magnetometry. The device is composed of two microfluidic channels formed between a glass cover slip (1 in x 1 in wide, 150 μm thick) and a molded block of polydimethylsiloxane (PDMS). The microfluidic channels are filled with a viscous fluid that contains target magnetic particles in a liquid suspension (see Methods). In addition to increasing the fluid viscosity, the fluid also serves to push the magnetic particles to the surfaces of the device (38, 39). We find that the magnetic particles are preferably confined under the top PDMS surface of the microfluidic channels. The intersection between the two channels forms the control chamber where particle manipulation and mapping of magnetic fields occurs.

We apply voltages at the four ends of the microfluidic channels and use feedback control in order to manipulate magnetic particles in the horizontal plane with nanoscale positioning accuracy (37-39). These voltages induce electroosmotic flow that moves the suspended magnetic particle toward desired directions by viscous drag (see Methods). To achieve vertical manipulation, we use magnetic actuation. We place a coil magnet under the microfluidic device that applies a magnetic field in the vertical direction. The magnetic field provides a downward force on the magnetic particle, and also orients its dipole moment along the vertical direction.

In order to perform local magnetometry, we deposit a dilute concentration of diamond nanocrystals (mean size of 25 nm) on the glass surface of the microfluidic device using a spin coating method. A microwave antenna (500 nm Au, 10 nm Ti) is lithographically patterned on the glass cover slip surface and drives the spin transition of the NV center to perform optically detected ESR measurements (7). Figure 1b shows an optical image of the microfluidic device with the integrated microwave antenna. We align the microfluidic channel such that the microwave antenna is as close as possible to the control chamber without directly penetrating into it. A tunable microwave signal generator (ROHDE&SCHWARZ SMC100A) drives the antenna in the frequency range, which spans at the ESR frequency of the NV center (2.87 GHZ).

The microfluidic device is mounted on a confocal microscope that excites the sample and collects emission (see Methods). Figure 1c shows an EM-CCD camera image of a single diamond nanocrystal and a nearby magnetic particle in the control chamber of the microfluidic device. For all experiments reported in this letter the magnetic particles are spherical maghemite magnetic beads with a radius of 500 nm (Chemicell, Germany). The diamond nanocrystal and the magnetic particle are highlighted with green and red boxes respectively. We select a diamond nanocrystal that is located close to the patterned microwave antenna in order to minimize the microwave power required to excite ESR transitions. Figure 1d shows an autocorrelation measurement taken from the diamond nanocrystal that exhibits clear anti-bunching, indicating that it contains only a single NV defect (see Methods).

**Manipulation of a Magnetic Nanoparticle in the Microfluidic Device**

We first investigate the ability to manipulate a magnetic particle within the microfluidic control chamber. Figure 2(a-c) shows a sequence of camera images of a single magnetic particle that we manipulate on the PDMS surface using feedback control to undergo a square spiral trajectory. Each panel shows the position of the particle at three different times (150, 300, and 450 sec). The cyan and magenta boxes indicate the start and stop positions of the trajectory. The white line shows the past history of the measured positions of the controlled particle.

In order to determine the accuracy of the particle manipulation by flow feedback control, we hold the particle at a desired location for 60 seconds and monitor its position in real time. We plot the measured positions in Figure 2d. Figure 2e,f show histograms of the measured positions along x and y coordinates respectively. From the standard deviation in the Gaussian fit (red solid line), we calculate a positioning accuracy of 48 (47) nm along the x (y) directions respectively.

To achieve motion of the magnetic particle in the direction orthogonal to the surface, we apply a current to the coil magnet under the device. The generated magnetic field applies a force that pulls the magnetic particle away from the PDMS surface and towards the glass surface. Figure 2g plots the measured distance of the magnetic particle to the bottom glass surface as a function of the current driving the coil magnetic. We measure this distance by sweeping the particle through the focus of the objective lens using the piezo stage that holds the microfluidic device. We determine the focused position from the minimum spot size and record the stage travel distance. We repeat the measurement five times for each value of the external current. The blue dots in Figure 2g show the values obtained for the five individual measurements, while the red line shows the average value of the data.

At each driving current, the magnetic particle stabilizes at a different distance relative to the bottom glass surface due to cancellation between fluid forces that push the particle upwards to the top PDMS surface, and magnetic gradient forces that pull it downwards towards the bottom glass surface. This interplay between chemical buoyancy versus magnetic downward force enables us to control the vertical position of the magnetic particle relative to the device surface. At 40 mA current, the particle is pulled almost all the way to the bottom glass surface. The applied magnetic field also serves to orient the dipole moment of the suspended magnetic particle along the direction of the applied field in the fluid, thereby eliminating the effect of tumbling of the particle that would otherwise randomize its orientation relative to the measurement direction of the NV center.

**Fluorescence Magnetometry of a Magnetic Particle**

In order to measure the magnetic field generated by the magnetic particle, we scan its position relative to the NV center and perform optically detected ESR measurements. We isolate a diamond nanocrystal that contains a single NV center on the glass surface in the control chamber. We apply a 50 mA current to the external coil magnet in order to pull the magnetic particle down to the glass surface where the NV center is located. Based on this current and the measurement results in Figure 2g, we estimate the center of the magnetic particle to be located 0.7 μm above the glass surface. Under these conditions, we scan the magnetic particles position relative to the NV center and obtain an ESR spectrum at each position.

We perform the ESR measurements using a digital lock-in approach, as originally proposed by Horowitz *et al* (26). This technique is insensitive to photoluminescence intensity fluctuations that arise from slow mechanical drift of the NV center relative to the tight Gaussian excitation laser due to the sample stage drift and vibration, as well as instability of the NV center emission caused by charge fluctuations on the surface (26, 40). We modulate the microwave power supply on and off at a 1 KHz rate and measure photoluminescence intensity synchronized with this modulation. We define the photoluminescence intensity contrast of the NV center as

$$C(f) = \frac{I_{on}(f) - I_{off}(f)}{I_{off}(f)}, \qquad (1)$$

where $I_{on}$ and $I_{off}$ are the photoluminescence intensity when the microwave is on and off respectively at each value of the microwave frequency.

Figure 3 shows the ESR spectrum of the NV center taken at three different positions of the magnetic particle. The left panels show the measured ESR spectrum, where blue dots represent measured data points and the red lines are Lorentzian fits to the measured data. The right panels show EM-CCD camera images that indicate the horizontal position of the magnetic particle (red box) and the NV center (green box). Panel a shows the case where the magnetic particle is 7.25 μm away from the NV center so that its contribution to the total magnetic field is negligible. The two ESR peaks correspond to the Zeeman split spin ground states of the NV center which are separated by 42.3 MHz. This large splitting is due to the applied magnetic field from the external coil magnet.

As we bring the magnetic particle to a distance of 3.62 μm (panel b), the Zeeman splitting reduces to 40.7 MHz because the field generated by the magnetic particle opposed the external field of the coil magnet. This change in Zeeman splitting provides a direct quantitative measure of the magnetic field given by the magnetic particle. At an even shorter distance of 1.50 μm (panel c), the Zeeman splitting is further reduced to 34.2 MHz. We note that in addition to the reduction in Zeeman splitting, the contrast of the ESR spectrum also reduces. We attribute this reduction in contrast to increased background from scattering of the white light source by the magnetic particle. As the magnetic particle gets closer to the diamond nanocrystal, it scatters more white light through the spatial filter, thereby increasing the background levels.

To map out the magnetic field distribution of the magnetic particle, we position it at various horizontal distances *r* relative to the NV center and measure the variation in Zeeman splitting. We can approximate the spherical magnetic particle as a magnetic dipole when its radius is small compared to the distance *r*. Figure 4a plots the calculated magnitude of the magnetic field along the glass surface for a magnetic dipole whose dipole moment points normal to the device surface. The open circles represent horizontal positions of the magnetic particle relative to the NV center where we sampled the magnetic field. Figure 4b plots the measured magnetic fields of the magnetic particle, obtained from the change in Zeeman splitting, as a function of distance *r*. Error bars denote the standard deviation of the Lorentzian fit used to find the center frequencies of the two Zeeman split ESR transitions. The red curve represents the theoretically predicted magnetic field along the glass surface for a point magnetic dipole. In these calculations, we treat amplitude of the particle magnetic moment and orientation of the NV center as fitting parameters (see Methods). We note that the dipole model is only accurate when the distance of the particle to the NV center exceeds the particle radius (500 nm). This condition is satisfied for virtually all the data points in the figure.

We estimate the shot-noise limited magnetic field sensitivity of our magnetometry system using the relation (7, 41)

$$\eta_B = 0.77 \frac{h}{g\mu_B} \frac{\Delta\nu}{C\sqrt{R}}, \quad (2)$$

where *h* is the Planck constant, *g* = 2 is the Lande g-factor for the NV center, and $\mu_B$ is the Bohr magneton. Using the data in Figure 3a, we calculate a maximum photoluminescence intensity contrast of *C* = 0.053, along with a hyperfine unresolved microwave ESR linewidth of $\Delta\nu$ = 7.2 MHz. The fluorescence count rate is measured to be *R* = 45 kCts/s. Inserting these values into Equation 2 we attain an estimated magnetic field sensitivity of 17.5 µT Hz$^{-1/2}$. This result compare well with previously reported magnetic field sensitivities of NV magnetometry demonstrated in a liquid environment (26, 27).

**Conclusions**

We have developed a microfluidic platform to perform localized NV magnetometry. We demonstrated three dimensional manipulations of single magnetic particle and performed magnetometry with high field sensitivity and with nanoscale spatial precision using a single NV color center. The results reported here using a single NV center can be directly extended to multiple NV centers at different locations of the device surface that serve as parallel sensors for vector magnetometry. Our results can also be adapted to exploit the electric field (42) and thermal gradient (43) sensing capabilities of NV centers to measure other physical parameters. The presented NV magnetometry system could ultimately lead to highly functional microfluidic systems that combine single particle manipulation and real-time sensing to enable a broad range of applications in the study of biological and chemical systems.

## Methods

### Fluid composition
The fluid solution is composed of 1.5 wt % rheology modifier (Acrysol RM-825, Rohm and Haas Co.) and 2.5 % by volume ethoxylated-15 trimethylolpropane triacrylate resin (SR-9035, Sartomer) in deionized water.  The rheology modifier is used to increase viscosity of the fluid to reduce Brownian motion of the magnetic particles, and the triacrylate resin creates a fluid chemistry that pushes the magnetic particles to the top PDMS surface of the microfluidic channels.

### Optical measurement setup
We mount the microfluidic device on an inverted confocal microscope system that excites and collects emission from a NV center and images a manipulated magnetic particle through the bottom glass cover slip using a 1.45 numerical aperture oil immersion objective.  A piezo stage attached to the microscope enables precise translation of the microfluidic device in three dimensions.  We excite the NV center using a 532 nm continuous wave laser and use a half-waveplate to rotate the polarization to achieve maximum fluorescence intensity.  In addition to the excitation laser, we use a white light source to image the magnetic particle.  The excitation laser beam is tightly focused onto the NV center while the white light beam is widely focused to a 20 μm diameter.

A beam-splitter sends 25% of the collected light to an EM-CCD camera (Hamamatsu C9100-13) to track the magnetic particle using a 10 Hz camera frame rate.  The remaining 75% of the collected light is spatially filtered by a pinhole aperture and is spectrally filtered using a 600-750 nm bandpass filter.  This combination of spatial and spectral filtering isolates the emission from the NV center and rejects the background signal from the scattered white light with a high extinction ratio.  The filtered NV emission is directed to an avalanche photodiode (PerkinElmer SPCM-AQR) to perform optically detected ESR measurements.  The signal can also be directed to a Habury-Brown Twiss (HBT) intensity interferometer composed of a 50/50 beam-splitter and two avalanche photodiodes, to perform second order correlation measurements.  A time interval analyzer (PicoQuant PicoHarp 300) performs time-resolved coincidence detection using the outputs of the two photon counters.

### Feedback flow control system
Details of the microfluidic control system have been reported in (37-39). The EM-CCD camera continuously monitors the position of a selected magnetic particle to nanoscale accuracy using sub-pixel averaging.  The feedback control algorithm compares the current position of the particle with the desired position and applies voltages at the four ends of the microfluidic channels using platinum electrodes.  These voltages then induce a corrective electroosmotic flow that moves the suspended magnetic nanoparticle from its current position to its desired position by viscous drag.  This process continuously moves the particle as desired with nanoscale accuracy.  This method enables two-dimensional control of a suspended magnetic particle in the microfluidic channel in real time.

**Fitting measured magnetic field to theoretical dipole model**

We assume the magnetic particle behaves like a magnetic dipole whose magnetic field is given by

$$\mathbf{B}_{dip} = \frac{\mu_0}{4\pi} \frac{3\mathbf{r}(\mathbf{m}\cdot\mathbf{r}) - \mathbf{m}r^2}{r^5}, \qquad (3)$$

where $\mu_0$ is the vacuum permeability and the magnetic moment of the dipole **m** points orthogonal to the microfluidic device surface. The magnetic field measured by the NV center is the projection of the magnetic field of the dipole $\mathbf{B}_{dip}$ onto the NV orientation $\boldsymbol{\beta}$ :

$$B_{NV} = \boldsymbol{\beta} \cdot \mathbf{B}_{dip} . \qquad (4)$$

We fit $B_{NV}$ to the measured data points using **m** and $\boldsymbol{\beta}$ as fitting parameters.

**Acknowledgements**

The authors would like to acknowledge financial support from the Physics Frontier Center at the Joint Quantum Institute.

**References**

1. Kurtsiefer C, Mayer S, Zarda P, & Weinfurter H (2000) Stable Solid-State Source of Single Photons. *Physical Review Letters* 85(2):290-293.
2. Beveratos A*, et al.* (2002) Single Photon Quantum Cryptography. *Physical Review Letters* 89(18):187901.
3. Jelezko F, Gaebel T, Popa I, Gruber A, & Wrachtrup J (2004) Observation of Coherent Oscillations in a Single Electron Spin. *Physical Review Letters* 92(7):076401.
4. Balasubramanian G*, et al.* (2009) Ultralong spin coherence time in isotopically engineered diamond. *Nat Mater* 8(5):383-387.
5. Gruber A*, et al.* (1997) Scanning Confocal Optical Microscopy and Magnetic Resonance on Single Defect Centers. *Science* 276(5321):2012-2014.
6. Manson NB, Harrison JP, & Sellars MJ (2006) Nitrogen-vacancy center in diamond: Model of the electronic structure and associated dynamics. *Physical Review B* 74(10):104303.
7. Taylor JM*, et al.* (2008) High-sensitivity diamond magnetometer with nanoscale resolution. *Nat Phys* 4(10):810-816.
8. Maze JR*, et al.* (2008) Nanoscale magnetic sensing with an individual electronic spin in diamond. *Nature* 455(7213):644-647.
9. Balasubramanian G*, et al.* (2008) Nanoscale imaging magnetometry with diamond spins under ambient conditions. *Nature* 455(7213):648-651.
10. Maertz BJ, Wijnheijmer AP, Fuchs GD, Nowakowski ME, & Awschalom DD (2010) Vector magnetic field microscopy using nitrogen vacancy centers in diamond. *Applied Physics Letters* 96(9):-.
11. Schoenfeld RS & Harneit W (2011) Real Time Magnetic Field Sensing and Imaging Using a Single Spin in Diamond. *Physical Review Letters* 106(3):030802.
12. Arcizet O*, et al.* (2011) A single nitrogen-vacancy defect coupled to a nanomechanical oscillator. *Nat Phys* 7(11):879-883.

13. Maletinsky P, et al. (2012) A robust scanning diamond sensor for nanoscale imaging with single nitrogen-vacancy centres. *Nat Nano* 7(5):320-324.
14. Rondin L, et al. (2012) Nanoscale magnetic field mapping with a single spin scanning probe magnetometer. *Applied Physics Letters* 100(15):-.
15. S Prawer and P Mulvaney and F Jelezko and J Wrachtrup and R E Scholten and L C L Hollenberg LPMaLTHaASaDASaCDHaJHCaKGaBCGa (2013) Ambient nanoscale sensing with single spins using quantum decoherence. *New Journal of Physics* 15(7):073042.
16. Ermakova A, et al. (2013) Detection of a Few Metallo-Protein Molecules Using Color Centers in Nanodiamonds. *Nano Letters* 13(7):3305-3309.
17. Kaufmann S, et al. (2013) Detection of atomic spin labels in a lipid bilayer using a single-spin nanodiamond probe. *Proceedings of the National Academy of Sciences* 110(27):10894-10898.
18. Grinolds MS, et al. (2013) Nanoscale magnetic imaging of a single electron spin under ambient conditions. *Nat Phys* 9(4):215-219.
19. Rondin L, et al. (2013) Stray-field imaging of magnetic vortices with a single diamond spin. *Nat Commun* 4.
20. Fu C-C, et al. (2007) Characterization and application of single fluorescent nanodiamonds as cellular biomarkers. *Proceedings of the National Academy of Sciences* 104(3):727-732.
21. Liu K-K, Cheng C-L, Chang C-C, & Chao J-I (2007) Biocompatible and detectable carboxylated nanodiamond on human cell. *Nanotechnology* 18(32):325102-325102.
22. McGuinness LP, et al. (2011) Quantum measurement and orientation tracking of fluorescent nanodiamonds inside living cells. *Nat Nano* 6(6):358-363.
23. Le Sage D, et al. (2013) Optical magnetic imaging of living cells. *Nature* 496(7446):486-489.
24. Hall LT, et al. (2012) High spatial and temporal resolution wide-field imaging of neuron activity using quantum NV-diamond. *Sci. Rep.* 2.
25. Kucsko G, et al. (2013) Nanometre-scale thermometry in a living cell. *Nature* 500(7460):54-58.
26. Horowitz VR, Alemán BJ, Christle DJ, Cleland AN, & Awschalom DD (2012) Electron spin resonance of nitrogen-vacancy centers in optically trapped nanodiamonds. *Proceedings of the National Academy of Sciences* 109(34):13493-13497.
27. Geiselmann M, et al. (2013) Three-dimensional optical manipulation of a single electron spin. *Nat Nano* 8(3):175-179.
28. Yi C, Li C-W, Ji S, & Yang M (2006) Microfluidics technology for manipulation and analysis of biological cells. *Analytica Chimica Acta* 560(1–2):1-23.
29. Hoyoung Yun and Kisoo Kim and Won Gu L (2013) Cell manipulation in microfluidics. *Biofabrication* 5(2):022001.
30. Steinert S, et al. (2013) Magnetic spin imaging under ambient conditions with sub-cellular resolution. *Nat Commun* 4:1607.
31. Micheal A, Satej C, Roland P, Shawn W, & Benjamin S (2005) Control of microfluidic systems: two examples, results, and challenges. *International Journal of Robust and Nonlinear Control* 15(16):785-803.
32. Cohen AE (2005) Control of Nanoparticles with Arbitrary Two-Dimensional Force Fields. *Physical Review Letters* 94(11):118102.
33. Chaudhary S & Shapiro B (2006) Arbitrary steering of multiple particles independently in an electro-osmotically driven microfluidic system. *Control Systems Technology, IEEE Transactions on* 14(4):669-680.
34. Komaee A & Shapiro B (2012) Steering a Ferromagnetic Particle by Optimal Magnetic Feedback Control. *Control Systems Technology, IEEE Transactions on* 20(4):1011-1024.

35. Nacev A, *et al.* (2012) Towards Control of Magnetic Fluids in Patients: Directing Therapeutic Nanoparticles to Disease Locations. *Control Systems, IEEE* 32(3):32-74.
36. Zhaolong S, Chen K, & Shapiro B (2012) Measuring low concentrations of fluorescent magnetic nanoparticles by fluorescence microscopy. *Manipulation, Manufacturing and Measurement on the Nanoscale (3M-NANO), 2012 International Conference on*, pp 283-287.
37. Ropp C, *et al.* (2010) Manipulating Quantum Dots to Nanometer Precision by Control of Flow. *Nano Letters* 10(7):2525-2530.
38. Ropp C, *et al.* (2010) Positioning and Immobilization of Individual Quantum Dots with Nanoscale Precision. *Nano Letters* 10(11):4673-4679.
39. Ropp C, *et al.* (2013) Nanoscale imaging and spontaneous emission control with a single nano-positioned quantum dot. *Nat Commun* 4:1447.
40. Bradac C, *et al.* (2010) Observation and control of blinking nitrogen-vacancy centres in discrete nanodiamonds. *Nat Nano* 5(5):345-349.
41. Dréau A, *et al.* (2011) Avoiding power broadening in optically detected magnetic resonance of single NV defects for enhanced dc magnetic field sensitivity. *Physical Review B* 84(19):195204.
42. Dolde F, *et al.* (2011) Electric-field sensing using single diamond spins. *Nat Phys* 7(6):459-463.
43. Toyli DM, de las Casas CF, Christle DJ, Dobrovitski VV, & Awschalom DD (2013) Fluorescence thermometry enhanced by the quantum coherence of single spins in diamond. *Proceedings of the National Academy of Sciences* 110(21):8417-8421.

**Figure Legends**

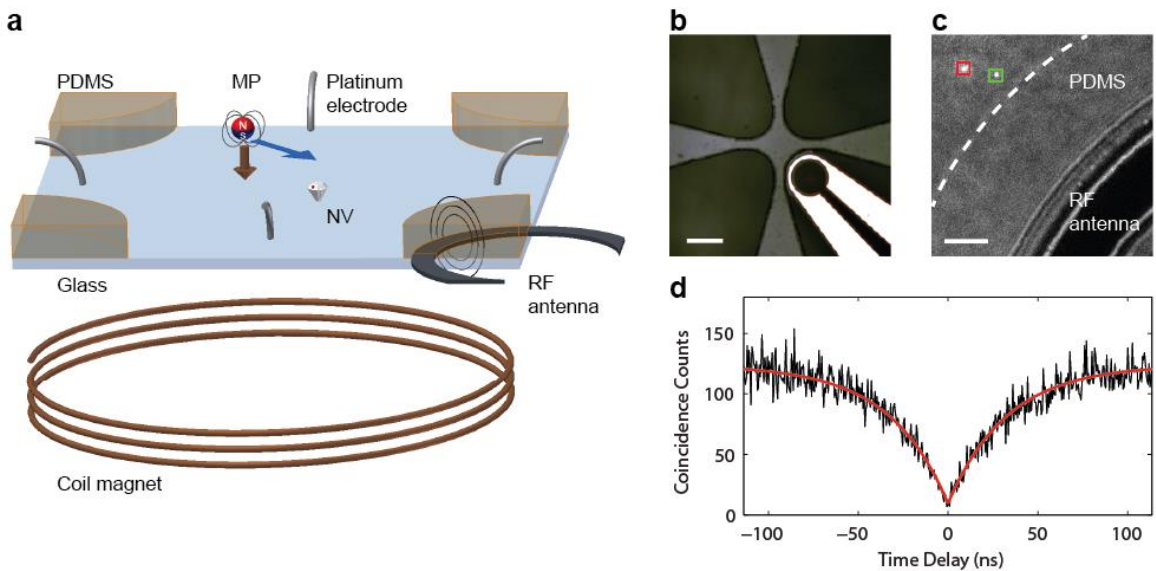

Figure 1. (a) Schematic of microfluidic device used to implement NV magnetometry. The magnetic particle is labeled by "MP". (b) Optical image of the control chamber of the microfluidic device along with integrated microwave antenna. Scale bar is 100 µm. (c) EM-CCD image of an immobilized NV center (green) and a suspended magnetic particle (red) inside the microfluidic control chamber. The white dashed line represents the boundary between the fluid and the PDMS sidewall of the channel. Scale bar is 5 µm. (d) Second order correlation measurement of a single NV center in a diamond nanocrystal.

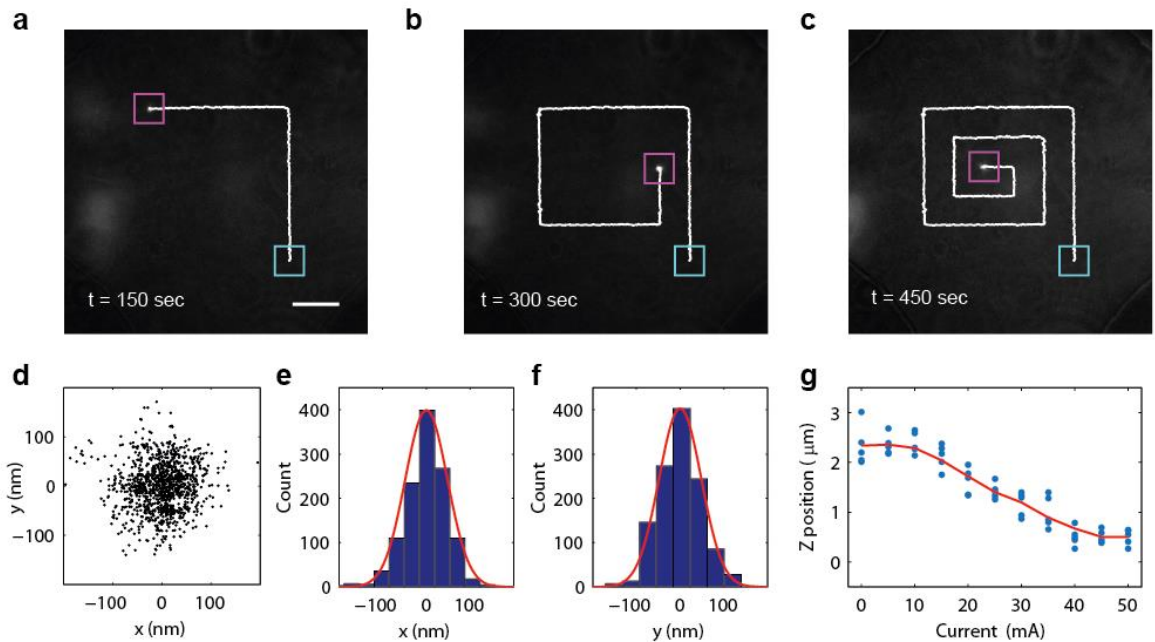

Figure 2. (a-c) Camera images of a magnetic particle manipulated along a 2D square spiral trajectory. The cyan box indicates the start position of the selected magnetic particle, and the magenta boxes indicate the stop location of the particle at 150 s, 300 s, and 450 s respectively. The white line shows the past history of the measured positions. Scale bar is 10 µm. (d) Scatter plots of the measured positions of the magnetic particle held in place using flow control for 60 sec. (e,f) Position histograms along the x and y axes of panel d. The solid lines are Gaussian fits to the position histograms. The standard deviation is 48 (47) nm along the x (y) directions respectively. (g) Vertical distance of the magnetic particle relative to the bottom glass surface as a function of the current driving the coil magnetic underneath the device. Blue dots show individually measured values while the red line shows the averaged values.

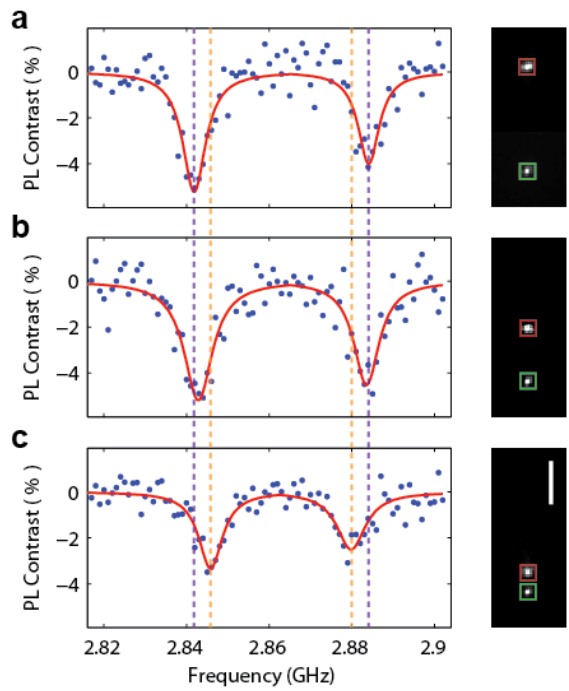

Figure 3. Optically detected ESR spectra (left) and camera images (right) of the magnetic particle (red square) and NV emission (green square) when the particle is at a distance of (a) 7.25 μm, (b) 3.62 μm, (c) and 1.50 μm from the NV center.  Scale bar is 3 μm.  The red lines in the spectra are Lorentzian fits to measurement data. Dashed lines in purple and orange are the center frequencies of the Lorentzian fits in (a) and (c).

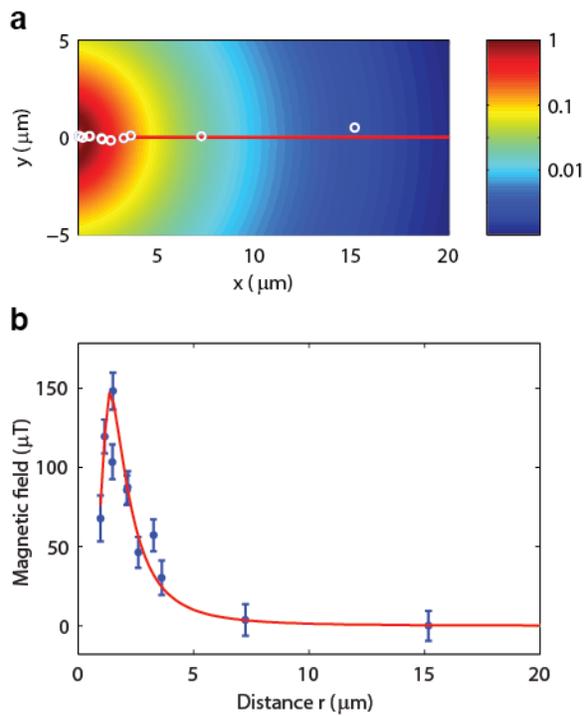

Figure 4. (a) Calculated magnetic field intensity of a magnetic dipole as a function of distance. Open white circles represent experimentally sampled positions of the magnetic field. (b) Measured magnetic field at the various positions represented by the open circles in panel a. Error bars denote the standard deviation of the Lorentzian fit to the ESR spectra. The red curve represents the theoretically calculated magnetic field for the magnetic particle, assumed to be a magnetic dipole.